\documentclass[12pt]{article}
\usepackage{arxiv}
\usepackage{graphicx} 
\usepackage{tikz}
\usepackage{url}
\usepackage{booktabs}
\usepackage{subcaption}
\usepackage{adjustbox}
\usepackage{multirow,multicol}

\usepackage{amsmath,amsfonts}
\usepackage{amsthm}

\usepackage{pifont}
\usepackage{makecell}

\title{Autobidding Arena: unified evaluation of the classical and RL-based autobidding algorithms}

\newcommand{\shorttitle}{{\small Autobidding Arena}}

\author{Andrey Pudovikov\thanks{Corresponding author}\\
MSU IAI, Moscow, Russia \\
\url{a.pudovikov@iai.msu.ru}\\
\And 
Alexandra Khirianova \\
MSU IAI \\
Moscow, Russia
\And 
Ekaterina Solodneva\\
MSU IAI \\
Moscow, Russia\\
\And
Aleksandr Katrutsa\\
MSU IAI \\
Moscow, Russia\\
\And
Egor Samosvat\\
Independent researcher\\
Moscow, Russia \\
\And
Yuriy Dorn \\
MSU IAI \\
Moscow, Russia\\
}

\begin{document}
\maketitle


\begin{abstract}
Advertisement auctions play a crucial role in revenue generation for e-commerce companies.
To make the bidding procedure scalable to thousands of auctions, the automatic bidding (autobidding) algorithms are actively developed in the industry. 
Therefore, the fair and reproducible evaluation of autobidding algorithms is an important problem. 
We present a standardized and transparent evaluation protocol for comparing classical and reinforcement learning (RL) autobidding algorithms.
We consider the most efficient autobidding algorithms from different classes, e.g., ones based on the controllers, RL, optimal formulas, etc., and benchmark them in the bidding environment.
We utilize the most recent open-source environment developed in the industry, which accurately emulates the bidding process.
Our work demonstrates the most promising use cases for the considered autobidding algorithms, highlights their surprising drawbacks, and evaluates them according to multiple metrics.
We select the evaluation metrics that illustrate the performance of the autobidding algorithms, the corresponding costs, and track the budget pacing.
Such a choice of metrics makes our results applicable to the broad range of platforms where autobidding is effective.
The presented comparison results help practitioners to evaluate the candidate autobidding algorithms from different perspectives and select ones that are efficient according to their companies' targets.
\end{abstract}


\section{Introduction}


E-commerce platforms actively use auction-based techniques to rank items based on user queries. 
In response to the user's request, a set of the most relevant ads is selected for display together with the relevant items. 
Such ads participate in the auction for paid slots, e.g., some top positions, in the user's query results.  
Each seller, the owner of such an ad, makes a bid for a particular slot, and the winner is the one whose ad is shown in this slot. 
The winner pays immediately, or after a specific event occurs, e.g., a user clicks on the shown ad~\cite{yin2016ranking, roy2022users}.  
Thus, bidding algorithms aim to provide the maximum number of target events for the least cost.



On major platforms, the number of queries and auctions can reach millions per day~\cite{roy2022users, khirianova2025bat}.
Since the auction depends on the user, query, and concurrent ads, every auction requires a particular bid from the seller.
Setting bids based on expert observations and experience, without algorithmic optimization, also known as manual bidding, became irrelevant for large platforms~\cite{wu2015predicting}. 


In contrast, modern platforms deploy automatic bidding (autobidding) systems that calculate bids that balance cost and value of winning the auction according to algorithms~\cite{lee2013real}. 
These algorithms not only consider the features of the seller, ad, user, and query, but can also incorporate the seller's requirements in generating bids.
For example, based on the seller's preferences, an algorithm can force the uniform budget spending across the target period, or set an upper bound on the average cost per click~\cite{yang2019bid}.

Many authors illustrate the benefits of their novel autobidding algorithm through comparisons with several baselines~\cite{lin2016combining, zhang2017managing, yang2019bid, stram2024mystique, khirianova2025bat, zhang2025adapting, BiCB}.
Such benchmarking is often made according to the specifically designed metrics, which are difficult to transfer to other domains or settings. 
Moreover, the recent advances in RL-based autobidding algorithms~\cite{wang2022roiCBRL,mou2022sustainableSORL} raise a natural question of whether they do outperform the \emph{tuned} classical methods. 
To answer this question, we perform a comparison of a broad range of classical and RL-based autobidding algorithms in a unified setup, using \emph{transparent and intuitive} evaluation metrics.
These metrics can be applied in various domains and demonstrate the core performance of autobidding algorithms, e.g., average number of clicks or conversions, the corresponding costs, e.g., cost per click, and the budget pacing schedule. 
To structure the comparison of the selected autobidding algorithms, we have developed the \emph{Autobidding Arena} framework.

Since most autobidding algorithms depend on multiple hyperparameters~\cite{yang2019bid, stram2024mystique,2023survey}, they are tuned to optimize a predefined metric using historical data, and then the algorithm is deployed with the tuned hyperparameters.
However, this pipeline ignores the impact of the tuned hyperparameters on other metrics, which can also be essential for the entire platform.
We present the results of extensive experiments that demonstrate how the hyperparameters tuned for the target metric affect the algorithms' performance in terms of the other metrics.
In our evaluation, the target metrics are the total number of clicks and conversions, the average auction win rate, and the uniformity of the budget pacing.

To perform the comparison in the unified setup that closely resembles the real-world bidding process, we utilize the recently presented environment~\cite{alibaba2024neurips} from a large e-commerce platform.
The key feature of this environment is that it generates the data that can be used as input to both classical and RL-based algorithms.
Note that since we have developed the Autobidding Arena framework for benchmarking algorithms in a modular manner, other environments can be further incorporated into it. 

The main contributions of our survey are the following:
\begin{enumerate}
    \item We develop the Autobidding Arena framework, which provides transparent and domain-free evaluation metrics.
    \item We demonstrate the performance of the selected classical and RL-based autobidding algorithms in an open-source simulation environment according to evaluation metrics.
    \item We empirically confirm that tuning hyperparameters in autobidding algorithms based on a predefined target metric can result in a decline in other metrics.      
\end{enumerate}

\section{Related works}

\paragraph{Surveys.}
There are two main surveys devoted to real-time bidding (RTB) and autobidding. 
The study~\cite{2023survey} provides a comprehensive overview of the RTB system, examining its components, including the role of click-through rate predictions. 
In contrast,~\cite{aggarwal2024auto} focuses specifically on autobidding through the lens of auction theory, analyzing optimal bidding strategies and price-of-anarchy in both truthful and non-truthful auction settings. 
While~\cite{2023survey} focuses mainly on the practical perspective of autobidding algorithms, the study~\cite{aggarwal2024auto} emphasizes their theoretical foundations. 
However, these surveys do not evaluate algorithms in terms of performance metrics, which motivates the presented study (see Table~\ref{table:comparison}).
\begin{table}[!ht]
\centering
\caption{Our study considers autobidding algorithms from a different product-based metrics perspective, while other surveys focus on auction theory and real-time bidding (RTB).}
\begin{tabular}{@{}lcccc@{}}
\toprule
    Survey & \multicolumn{1}{c}{Autobidding} & \multicolumn{1}{c}{Auctions theory} & \multicolumn{1}{c}{RTB} & \multicolumn{1}{c}{Metrics} \\
    \midrule
    \cite{aggarwal2024auto} & \textcolor{green}{\checkmark} & \textcolor{green}{\checkmark}& \textcolor{red}{\texttimes} &\textcolor{red}{\texttimes} \\
    \cite{2023survey} & \textcolor{green}{\checkmark} &\textcolor{red}{\texttimes} &  \textcolor{green}{\checkmark}  &\textcolor{red}{\texttimes} \\
    Our & \textcolor{green}{\checkmark} &\textcolor{red}{\texttimes} &\textcolor{red}{\texttimes} & \textcolor{green}{\checkmark} \\
\bottomrule
\end{tabular}
\label{table:comparison} 
\end{table}

\paragraph{Environment.}
A significant bottleneck in the research of autobidding algorithms is the scarcity of open datasets. 
The community largely relies on a limited number of publicly available benchmarks. 
such as the iPinYou dataset~\cite{ipinyou}, which remains a foundational resource; the recent BAT dataset~\cite{khirianova2025bat}; and the Alibaba dataset~\cite{alibaba2024neurips}. 
Thus, the majority of studies utilize the same few public sources and create their modifications~\cite{zhang2014optimal, zhang2015statistical, zhang2017managing, chen2023coordinated, wu2018budgetDRLB}.
Alternatively, researchers resort to using synthetic data generated from their proprietary, closed-data logs \cite{xu2015smart, zhang2015statistical,geyik2016joint, zhang2017managing,stram2024mystique} (see more in Section~\ref{sec:environment}).

\paragraph{Metrics.}
Since the autobidding algorithms crucially impact of the e-commerce platforms, along with classical metrics~\cite{zhang2014optimal, cai2017real,yang2019bid,gligorijevic2020bid,zhou2021efficient}, there are many highly specialized or platform-specific metrics.
These metrics are needed for internal and external production diagnostics~\cite{lucier2024autobidders,wang2022roi,kong2022not,balseiro2021robust,zhu2017optimized}.
The internal diagnostic aims to correct the algorithm's operation, while the external one evaluates the algorithm's performance.
Exploiting these metrics complicates straightforward comparison of algorithms. 
Therefore, we propose criteria for selecting the most transparent and universal metrics.

\paragraph{Algorithms.}
The first autobidding algorithms were na\"ive heuristics~\cite{zhang2014optimal, zhang2017managing, kitts2017ad} based on simple static rules (e.g., a fixed constant bid or a linear function of predicted click probability). 
They were followed by controller-based systems (PID), which addressed the problem of adhering to constraints through feedback loops and the dynamic adjustment of bids~\cite{zhang2016feedback,yang2019bid}. 
Furthermore, LP models reformulate the autobidding problem as a constrained optimization task and yield an optimal bidding formula~\cite{balseiro2015repeated,lucier2024autobidders}. 
Finally, breakthroughs in machine learning have led to hybrid approaches that combine predictive ML~\cite{BiCB} and RL methods~\cite{wu2018budgetDRLB, cai2017realRLB, USCB}.

Autobidding algorithms are also distinguished by their optimization objectives~\cite{2023survey}. 
Some of them maximize the cost-effective budget spending for clicks or conversions~\cite{zhang2014optimal, zhang2017managing, kitts2017ad, wu2018budgetDRLB}. 
Others prioritize smooth budget pacing~\cite{chen2023coordinated,stram2024mystique}.
Furthermore, certain algorithms are engineered to maximize platform-side metrics, such as revenue, market efficiency, or welfare~\cite{balseiro2015repeated, chen2023coordinated, lucier2024autobidders, su2024a}.

\section{Environment for autobidding simulations}
\label{sec:environment}
Autobidding algorithms are a key component of the complex autobidding systems that operate on e-commerce platforms.
The standard approach to confirm the gain from deploying novel algorithms is to perform an A/B test~\cite{li2022vulnerabilities, wen2022cooperative,gao2025generative}.
Although this approach can demonstrate the performance of the novel algorithm in a production setup, it requires much effort for proper design and could degrade the seller's experience during testing~\cite{zhang2016bidaware}.
Therefore, to select the most promising candidates for A/B tests, a properly designed simulation environment is used for offline benchmarking of the broad range of autobidding algorithms~\cite{Jeunen2022,alibaba2024neurips}. 

We have developed the environment based on the dataset from the Auto-Bidding in Large-Scale Auctions challenge~\cite{alibaba2024neurips} released by a large e-commerce company in the competition track of NeurIPS 2024.
Our environment manages the interaction with sellers similarly to the previous works~\cite{zhang2014optimal, geyik2016joint, lin2016combining,  zhang2016feedback, yang2019bid, zhou2021efficient, he2021unified, chen2023coordinated, 2023survey, stram2024mystique, lucier2024autobidders, khirianova2025bat}.
We consider $S=48$ independent sellers and two independent time periods, each split into $T=48$ timestamps $t=1,\ldots,T$ such that the first period is used for tuning algorithms and the second one is used for validation. 
In every timestamp, the environment simulates thousands of second-price auctions that require bids from the seller.
Every seller~$s$ corresponds to a single item and uses the same autobidding algorithm in auctions independently of other sellers.

For seller $s$ and auction $a$, the algorithm can take as input the historical data of previous auction outcomes, seller's budget $B_s^{(0)}$, upper bounds on costs per clicks $CPC$, and conversions $CPA$.
In addition, to improve bidding performance, the algorithm can use a probability of click if seller~$s$ has won the auction~$a$ denoted as $CTR_{as} = \mathbb{P}(click \mid win)$ and a probability of conversion if a user clicks on item $\mathbb{P}(conversion \mid click)$ denoted as $CVR_{as}$.
These quantities are estimated from the given data in the original dataset according to the procedure described in Appendix \ref{appendix:CTRCVR}.

A seller aims to submit a bid sufficiently large to win in the second-price auction.
In the case of winning the auction, the item is displayed to a user and may potentially result in a click or conversion.
Each seller participates in its own set of auctions, which do not intersect with those of other sellers.

The seller $s$ submits his own bid to the environment, and the environment compares the submitted bid with known bids from the available internal auction logs.
Then the environment responds to the seller to indicate whether the auction has been won or not, and if the seller has won, which cost has been paid for the dislplay.
In addition, if the seller wins the auction, the environment reports the resulting click $Clicks_{as} \sim Bernoulli(CTR_{as})$ and conversion $Cnv_{as} \sim Bernoulli(CTR_{as} \cdot CVR_{as})$. 

\begin{figure}[!ht]
    \centering
\includegraphics[width=0.5\linewidth]{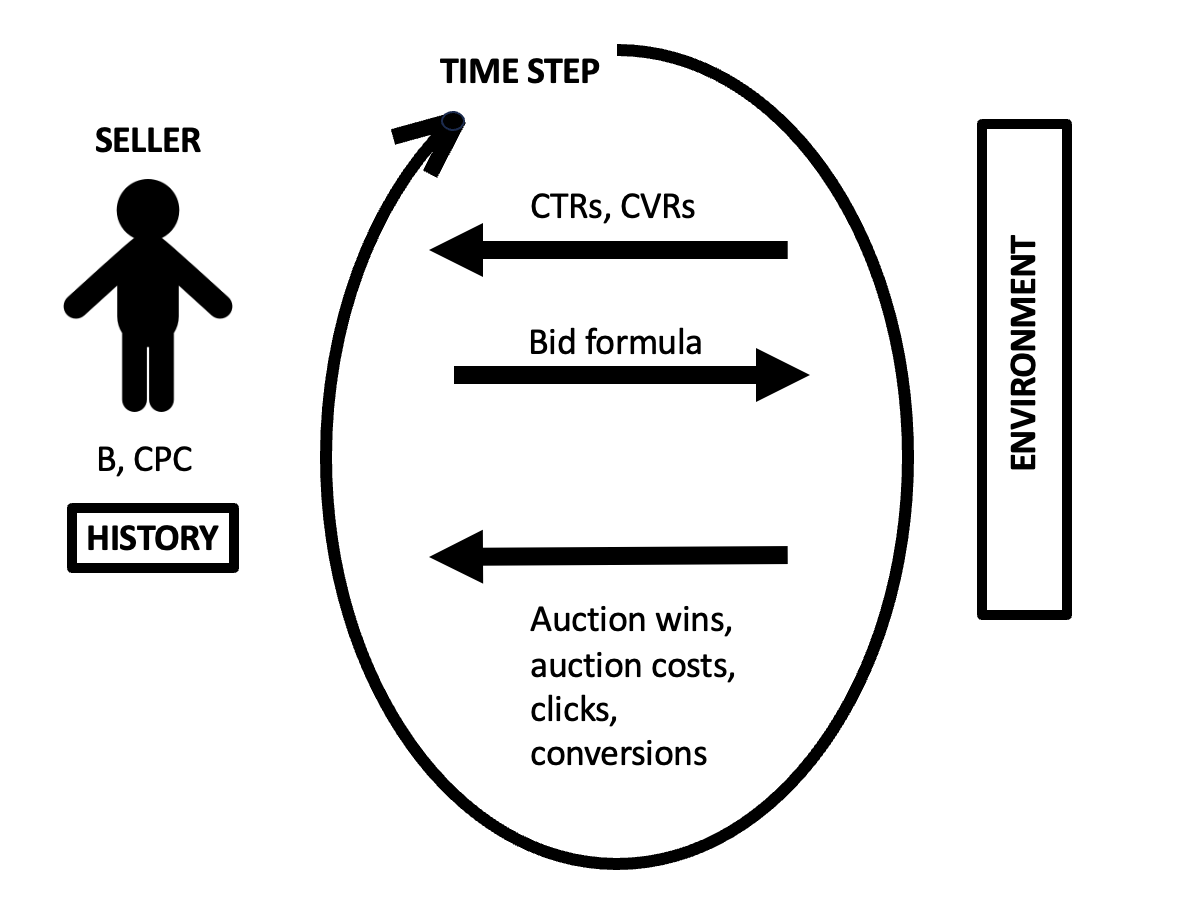}
    \caption{The pipeline of interaction between the seller and the environment at each timestamp. The seller sets an available budget, $B$, and an upper bound on the cost-per-click, $CPC$. Then, the environment reports CTR and CVR values, and the autobidding algorithm on the seller side produces bids. Based on the received bids, the environment evaluates the auction result and its outcome, including clicks and conversions.}
    \label{fig:environment}
\end{figure}

After that, the historical data for seller $s$ is updated with recent auction outcomes, and the autobidding algorithm can be adjusted according to the internal procedure.
Note that our environment provides an auction summary to sellers; however, the logged winning price in the environment remains the same.
The summary of the developed environment is presented in Figure~\ref{fig:environment}.

In the bidding process, the purpose of the autobidding algorithm is to estimate bids that enable the seller to achieve the maximum number of ad displays, clicks, and/or conversions while satisfying the stated cost or budget pacing constraints.
Formally, denote by $x_{as}$ an indicator of the winning an action $a$ by seller $s$, i.e. 
\[x_{as} = 
\begin{cases}
1, & \text{if seller $s$ wins auction $a$}, \\
0, & \text{otherwise}.
\end{cases}
\]
Similarly, we recap the previously introduced binary indicators $Click_{as}$ and $Cnv_{as}$.
Here, $Click_{as} = 1$ if the ad is shown ($x_{as} = 1$) and a click occurs; otherwise, it is $0$.
Furthermore, $Cnv_{as} = 1$ if a click occurred ($Click_{as} = 1$) and that click led to a conversion; otherwise, it is $0$.
In addition, we can estimate the cost of winning the auction $a$ by seller $s$ as
\[
Cost_{as} = x_{as} \cdot wp_{a},
\]
where $wp_a$ is the winning price from the dataset for an auction~$a$.
Based on the introduced notations, we can formally derive the constraints on the costs.
In particular, the budget constraint for the seller $s$ can be naturally expressed in the form:
\begin{equation}
\sum_a^{A_s} Cost_{as} \leq B_s^{(0)},
\label{eq:b_constr}
\end{equation}
where $A_s$ is the total number of auctions in which seller $s$ is participated, and $B_s^{(0)}$ denotes the starting budget at $t=0$.
Cost-per-click and cost-per-action constraints can be formalized as
\begin{equation}
eCPC \leq CPC, \quad eCPA \leq CPA,
\label{eq::cost_constraints}
\end{equation}
where $CPC$ and $CPA$ are pre-defined upper bounds on the corresponding costs, while the left-hand sides $eCPC$ and $eCPA$ are expected values of corresponding costs and computed as follows:
\begin{equation}
    eCPC = \frac{\sum_{s} \sum_{a}^{A_s} Cost_{as}}{\sum_{s} \sum_a^{A_s}  Click_{as}}
    \label{eq::ecpc}
\end{equation}
\begin{equation}
    eCPA = \frac{\sum_{s} \sum_{a}^{A_s} Cost_{as}}{\sum_{s} \sum_{a}^{A_s} Cnv_{as}}.
    \label{eq::ecpa}
\end{equation}
We use $CPC = 0.5$ and $CPA = 0.05$ in our environment to ensure that the case of winning all auctions and receiving all clicks and conversions is impossible.
The available starting budget for each seller $B_s^{(0)}=10000$ can be spent over $T=48$ timestamps.

Thus, we have specified the design of the environment used in our Autobidding Arena framework and formally described the constraints that autobidding algorithms have to satisfy.
The next ingredient of our framework is a set of carefully selected evaluation metrics that provide a fair and transparent comparison of the autobidding algorithms.
We present these metrics and the criteria for their selection in the next section.

\section{Autobidding performance metrics}\label{sec::metrics}

Studies on autobidding algorithms typically compare the new algorithm with baselines using several metrics. 
Along with classical metrics, such as the number of clicks and conversions, authors often use custom and/or domain-specific metrics. 
The active exploitation of such metrics complicates the fair comparison of autobidding algorithms and the extension of the presented comparison results to baselines not presented in a particular paper. 

In contrast, to fairly compare the autobidding algorithms, we first specify the criteria for the evaluation metrics, second review the existing metrics in the literature, and finally select such metrics that satisfy the specified criteria.
In particular, we focus on the transparent, universal, and aligned with environment assumptions metrics.
\emph{Transparent metrics} are computed without non-intuitive normalizations or poorly motivated formulas.
\emph{Universal metrics} are independent of business- or production-specific concepts, such as including delivery in an order, following a link to an external resource, shopping cart addition volume, etc.
Finally, \emph{the metrics aligned with environment assumptions} are metrics applicable to the task of optimizing the bid per slot for a seller with access to the values provided by our environment.




A detailed description of each selection criterion, along with a comprehensive table cataloging popular metrics from the literature and their adherence to criteria, can be found in the Appendix~\ref{appendix:MetrEval}. 
Below, we provide a list of metrics methods selected for comparison.

\subsection{How well autobidding works?}

These are the metrics, which are widely used in evaluation of autobidding algorithms~\cite{zhang2014optimal, cai2017real,gligorijevic2020bid} and are a natural measure of achieving the direct goals of autobidding: winning auctions - and therefore, impressions to users based on their queries, clicks on displayed ads, and some action based on the click (call to the seller, deal, delivery).

The auction wins rate $AWR$ is computed  as
\begin{equation}
AWR = \frac{\sum_s \sum_a^{A_s} x_{as}}{\sum_s A_s}
\label{eq::awr}
\end{equation}
and is applicable in the case of one slot~\cite{gligorijevic2020bid}. 
Following the experiment's design, it can evaluate all algorithms.  
While some problem statements do not consider the case of further actions as a conversion to contact or to deal, it is also important~\cite{yang2019bid} to sellers. 

The number of clicks ($Clicks$) achieved from ad displays to users is computed as 
\begin{equation}
Clicks = \sum_{s} \sum_{a}^{A_s} Click_{as}.
\label{eq::clicks_sum}
\end{equation}
and evaluates the performance of the autobidding algorithms, as it indicates the revenue prospects of the seller and the platform's revenue.

The number of target user actions achieved after clicking on an ad indicates the success of the algorithm if there is a certain action that a user could perform after clicking on the ad. 
Examples of such actions include concluding a deal with the seller, making a purchase, ordering delivery, or subscribing. 
In our environment, the total number of such events is referred to as total conversions, denoted as $Cnv$ and computed according to the following equation:
\begin{equation}
Cnv =  \sum_{s} \sum_{a}^{A_s} Cnv_{as}.
\label{eq::conversion_metric}
\end{equation}

\subsection{How cost autobidding works? }
\label{subsec::metrics_liquidity}
Here, we provide metrics that are economically important for both the seller and the platform. 
They correspond to the average cost estimate for auction win,
click, or conversion. 
Such estimators are widely used in the literature both to estimate the liquidity of the algorithm and its derivatives (elasticity, fairness, etc.)~\cite{cai2017real,gligorijevic2020bid, zhou2021efficient}. 
In addition to the $eCPC$~(\ref{eq::ecpc}) and $eCPA$~(\ref{eq::ecpa}) metrics introduced in Section~\ref{sec:environment}, the expected cost per 1000 auction wins ($eCPM$) is used and computed as
\begin{equation}
eCPM =\frac{\sum_{s} \sum_a^{A_s} Cost_{as}}{\sum_s \sum_a^{A_s} x_{as}} \cdot 1000.
\label{eq::ecpm}
\end{equation}
The normalization of $eCPM$ per thousand auctions is a tribute to the established accounting of the average cost per thousand displays on major platforms~\cite{liu2022real, 2023survey, luo2024puros, zhang2025adapting}.

\subsection{How uniform is budget pacing?}
\label{subsec::metrics_budget}

The autobidding algorithms have strict limitations from the budget of each campaign \cite{zhang2016optimal,khirianova2025bat}.
The series of works assumes that the quality of the algorithms should be assessed in terms of uniformity of the seller's money spending~\cite{zhang2016optimal,zhang2016feedback,khirianova2025bat}. 
To evaluate the budget pacing, we use the following equation~\cite{khirianova2025bat}: 
\begin{equation}
RMSE = \sum_s\sqrt{\frac{\sum_{t} (B_{*}^{(t)} - B_s^{(t)})^2}{T}},
\label{eq::rmse}
\end{equation}
where $B_s^{(t)}$ is the current balance of seller $s$ for the timestep $t$, $B_{*}^{(t)} = \frac{B \cdot (T-t)}{T}$ is the target balance. 
We assume that the optimal budget expenditure is linear throughout the entire campaign lifetime.
The correlation between the uniform budget pacing and other metrics is analyzed in experiments. 

\section{Autobidding algorithms}
\label{sec::criteria}
In this section, we provide a summary of the considered classical and RL-based autobidding algorithms.
For comparison purposes, we select the autobidding algorithms that are 
\begin{itemize}
    \item clearly and completely described in the source papers, i.e., no undefined or hidden constants and functions used inside,   
    \item compatible with the  simulation environment, see Section~\ref{sec:environment},
    \item recent and attracted the most attention from the community
    \item separated from other ingredients of autobidding systems, like CTR predictions or winning price estimation 
    \item the most efficient among diverse representatives of the autobidding algorithms classes.
\end{itemize}

We evaluate the popular autobidding algorithms according to these criteria and summarize our findings in Appendix~\ref{app:algos}.
The following sections provide brief descriptions of the selected non-RL and RL autobidding algorithms used in our experiments. 

\subsection{Non-RL autobidding algorithms}
\label{sec::non-rl}

This section presents a brief survey of the non-RL autobidding algorithms selected for comparison in our Autobidding arena framework.
They are grouped into classes based on underlying ideas.
Denote by $bid_{\mathcal{A}}$ the bid given by algorithm~$\mathcal{A}$. 
Since the bidding appears in every auction independently, we skip the indices of auction $a$, seller $s$, and timestep~$(t)$ for readability. 

\paragraph{Heuristic autobidding algorithms}
These autobidding algorithms are designed based on simple heuristics or hypotheses that lead to analytical formulas for bidding.
For example, study~\cite{zhang2014optimal} considers the constant bidding
\begin{equation}
bid_{const} = bid_0,
\end{equation}
where $bid_0$ is a hyperparameter, and random bidding that samples a bid randomly from the pre-defined segment: 
\begin{equation}
bid_{rand} = \mathrm{Uniform}(bid_{\min}, bid_{\max}),
\end{equation}
where $bid_{\min}$ and $bid_{\max}$ are bounds on possible bids extracted from the historical data, independent for each seller.
The linear bidding~\cite{zhang2014optimal} assumes that the bid is proportional to the CTR estimate: 
\begin{equation}
bid_{lin} = \alpha \cdot CTR \cdot obj, \quad obj = \begin{cases}
    CVR, & \mathfrak{M}_{tar} \text{ is } Cnv~(\ref{eq::conversion_metric})\\ 
    1, & \text{otherwise}.
\end{cases}
\label{eq:linear}
\end{equation}
where $\alpha$ is a hyperparameter, too, and $\mathfrak{M}_{tar}$ denotes the target metric used to tune hyperparameter.
The CostMax heuristic~\cite{zhang2014optimal} sets the bid proportional to the maximum desired bound:
\begin{equation}
bid_{CostMax} =b \cdot CPX, \quad 
CPX = \begin{cases}
    CPA, & \mathfrak{M}_{tar} \text{ is } Cnv\\ 
    CPC, & \text{otherwise,}
\end{cases}
\end{equation}
where $b$ is a hyperparameter. 
Study~\cite{zhang2014optimal} considers the real-time bidding (RTB) problem and explicitly maximizes the total number of clicks through two heuristic approximations of the winning rate.
The first approximation leads to the following simple bidding formula
\begin{equation}
bid_{ORTB_1} = \sqrt{\frac{c}{\lambda} \cdot CTR \cdot obj + c^2} - c
\end{equation}
and the second one leads to a more complicated expression for bidding:  
\begin{equation}
\begin{split}
& bid_{ORTB_2} = c\left[d^{1/3} - \left(\frac{c}{\lambda \cdot d}\right)^{1/3}\right], \\
&d=\frac{CTR\cdot obj}{\lambda} + \sqrt{\left(\frac{CTR\cdot obj}{\lambda}\right)^2 + \frac{c^2}{\lambda^2}},
\end{split}
\end{equation}
where $c, \lambda$ are the hyperparameters and $obj$ is similar to~(\ref{eq:linear}).


Although the heuristic autobidding algorithms are non-optimal and straightforward, they can still be tuned and show promising results~\cite{he2021unified, 2023survey, khirianova2025bat}.
However, there is a separate branch of autobidding algorithms that is based on solid theoretical justification and the solution of optimization problems, which we discuss in the following paragraph.

\paragraph{Autobidding algorithms based on solving optimization problems}
An alternative approach to heuristic autobidding algorithms is to derive the bidding formula from the solution of the explicit optimization problem.
Study~\cite{yang2019bid} states the problem of maximization the total number of conversion  $\sum_a x_{as}  {CTR}_{as}  {CVR}_{as}$, where $x_{as} \in \{0,1\}$ is an indicator that seller $s$ wins auction $a$, subject to budget and cost per click constraints.
After relaxation of the binary constraints to $x_{as} \in [0, 1]$ and solving a dual problem to the relaxed one, the following equation for bid is derived:
\begin{equation}
   {bid}_{OPT} = \frac{CTR}{p+q} obj + \frac{q \cdot CTR}{p+q}  \cdot  CPC,
\label{eq:pid2019}
\end{equation}
where $p,q > 0$ are optimal dual variables corresponding to the budget and cost-per-click constraints. 
Incorporating additional constraints, e.g., CPM, CPA, etc, to the considered optimization problem is discussed in the study~\cite{he2021unified}.








However, estimating optimal values for dual variables $ p$ and $ q$ is computationally infeasible in an online setting, where they must be recomputed before every upcoming auction.
Therefore, there are multiple approaches to approximate $p$ and $q$ on the fly, such as predictions based on a separate model or incremental updates using an auxiliary optimization method, etc.






The latter approach is discussed in works~\cite{balseiro2015repeated,lucier2024autobidders}, which propose an iterative corrections procedure to satisfy budget and cost per click constraints through the adjusting parameter $\mu^{(t)}$ in the following bidding formula:
\begin{equation}
    bid_{BROI} = \frac{CTR \cdot obj}{1 + \mu^{(t)}}, 
    \label{eq::broi}
\end{equation}
where $obj$ is the same as in~(\ref{eq:linear}).
Note that, formula~(\ref{eq::broi}) became equivalent to linear bidding heuristic~(\ref{eq:linear}), if $\mu^{(t)}$ would not incrementally updated through the internal procedure. 

Thus, although autobidding algorithms based on solving optimization problems provide more reasonable and theoretically-based bidding formulas, they suffer from excessive computational complexity, which makes them impractical.
Surprisingly, controller functions developed within the control theory~\cite{aastrom2021feedback,bennett1993development} appear to be a reasonable trade-off between computational efficiency and the theoretical basis of the bidding formula.
We discuss controllers demonstrating the most promising results in the next paragraph.










\paragraph{Controllers for autobidding algorithms}
The incorporation of the controller framework into the autobidding algorithms reveals a trade-off between the non-adaptive fast algorithms based on different heuristics and fully adaptive ones based on solving large-scale optimization problems.



For example, Fb-control~\cite{zhang2016feedback} is a feedback control algorithm that adjusts bids based on the difference between the target and observed values of the metric. 
If the target metric corresponds to the uniform budget pacing~(\ref{eq::rmse}), then Fb-control looks as follows:  
\begin{equation}
\begin{split}
bid_{Fb} ^{(t)}  &= bid_{Fb}^{(t-1)}\exp(\phi_{Fb}^{(t)}) \\
e^{(t)}  & =B_{*} ^{(t)}-B^{(t)}  \\
e_g ^{(t)}  & =B^{(t)}-B^{(t-1)}-[B_{*}^{(t)}-B_{*}^{(t-1)}]  \\
\phi_{Fb}^{(t)}  &=\lambda_1 e^{(t)}+\lambda_2 \sum_{t_i=1}^t e^{(t_i)} +\lambda_3 e_g^{(t)},
\end{split}
\label{eq:pid}
\end{equation}
where $\phi_{Fb}^{(t)}$ is the bid adjustment at timestep $t$, $e^{(t)}$ is the error between target $B_{*}^{(t)}$ and observed $B^{(t)}$ values of budget at timestamp~$t$, and $\lambda_1, \lambda_2, \lambda_3$ are the tuned hyperparameters.
If another metric is primary, the equations for $e^{(t)}$ and~$e^{(t)}_g$ is updated, respectively.

The simplified version of Fb-control is Fb-control-WL~\cite{zhang2016feedback}, which takes into account only the memory term to smooth bid adjustments over time and error $e^{(t)}$:
\begin{equation}
\phi_{FbWL}^{(t)} =  \phi_{FbWL}^{(t-1)} + \lambda_4 e^{(t)},
\end{equation}
where $\lambda_4$ is a tuned hyperparameter and $e^{(t)}$ is similar to~(\ref{eq:pid}).

Another promising controller-based autobidding algorithm is Mystique~\cite{stram2024mystique}, which also focuses on the budget pacing.
It controls the signal for bids through deviations of the budget spending function $B^{(t)}$ and its derivative:
\begin{equation}
\phi_{Mstq}^{(t)} =  \phi^{(t-1)} + w_s e^{(t)} +w_g e_g^{(t)},
\end{equation}
where $w_s,w_g$ are the parameters calculated based on $e^{(t)},e_g^{(t)}$ to follow the desired budget pacing.
PID controller from~\cite{yang2019bid} applies controller paradigm to~(\ref{eq:pid2019}) and  sequentially tunes
$p,q$ based on~(\ref{eq:pid}), where the error terms are computed for estimating budget $B^{(t)}$ and $eCPC^{(t)}$.

M-PID from work \cite{yang2019bid} is a multivariable PID controller that takes into account the cross-interaction of both PID parameters:
\begin{equation}
\begin{pmatrix}
\phi_{pM}^{(t)} \\
\phi_{qM}^{(t)}
\end{pmatrix}
=
\begin{pmatrix}
\gamma_p & 1 - \gamma_p \\
1 - \gamma_q & \gamma_q
\end{pmatrix}
\begin{pmatrix}
\phi_p^{(t)} \\
\phi_q^{(t)}
\end{pmatrix}
\end{equation}
where $\gamma_p, \gamma_q$ are optimized cross-interaction parameters.

\subsection{RL autobidding algorithms}


Reinforcement learning (RL) naturally addresses the sequential decision-making nature of autobidding.
Deep learning-enhanced RL algorithms can effectively handle complex impression and advertiser features.
Additionally, flexible reward engineering enables multi-objective optimization of budget, ROI, and performance metrics while learning adaptive strategies from auction experience.

\subsubsection{Markov Decision Process for autobidding}

The RL autobidding algorithms are based on modeling the bidding process as a Markov Decision Process (MDP). 
Modern approaches often utilize an impression-level MDP framework~\cite{cai2017realRLB} that makes bidding for each ad display.
The alternative approach to MDP formulation~\cite{zhang2023personalizedPerBid, wang2017ladder} focuses on keyword data, which complements the standard data for autobidding algorithms and is not publicly available.
Therefore, we do not consider RL autobidding algorithms that rely on such data within our benchmark. 

In the autobidding context, an advertiser is represented by an agent operating within the auction environment by making bids. 
The autobidding MDP is defined by the tuple $(\mathcal{S}, \mathcal{A}, \mathcal{P}, r)$, where $\mathcal{S}$, $\mathcal{A}$ represent a state and action spaces, respectively, $\mathcal{P}$ is transition probability distribution of moving to state $\hat{s}$ given current state $s'$ and action $a$, and $r: \mathcal{S} \times \mathcal{A} \to \mathbb{R}$ is a reward function that helps agent to learn its policy $\pi$.
The agent policy $\pi: \mathcal{S} \to \mathcal{A}$ represents the sequence of agents' decisions. 
The selected examples of these components in RL-based autobidding algorithms are listed below.

\paragraph{State representation}

The basic approach to state design is represented in the CMDP~\cite{du2017improvingCMDP} framework, which employs only predicted CTR as state.
Other RL autobidding algorithms use the extended state representation.
For example, the RLB~\cite{cai2017realRLB} method defines states using current time, remaining budget, and impression features, providing the agent with essential campaign context.
Other methods further extend state representation by incorporating auction outcomes and delivery metrics. 
DRLB~\cite{wu2018budgetDRLB} includes current time, remaining budget, number of opportunities left, CTR, budget consumption rate, CPM, cost-per-impression, winning ratio, and ROI. 
Similarly, FAB~\cite{liu2020dynamicFAB} focuses on information from the last elapsed timestamp, tracking budget ratio, cost ratio, CTR, and win rate. 
USCB~\cite{USCB} incorporates campaign-level information such as remaining time, remaining budget, budget pacing, and cost-related (CPC, CPA) or non-cost-related (e.g., CTR) KPI ratios.
The most comprehensive approach is taken by CBRL~\cite{wang2022roiCBRL}, which combines impression-level features with aggregate performance metrics, market prices, and ROI dynamics. 
This helps agents to adapt to changing auction environments.
Thus, the state of the agent can include the following ingredients: impression features (e.g., user features, item position in the search results), advertiser features (e.g., item category, CTR), and auction dynamics (e.g., market prices).

\paragraph{Action space design}
Action spaces in RL autobidding algorithms typically represent either explicit bid values or scaling factors in heuristic bid formulas, e.g., in BROI~(\ref{eq::broi}) or linear~(\ref{eq:linear}) algorithms.
A bid generated by action affects the auction outcome for an agent and the next agent's state. 
The straightforward approach from RLB, CMDP, and CBRL~\cite{cai2017realRLB, du2017improvingCMDP, wang2022roiCBRL} designs the action space as possible bid values and allows agents to output a bid for each auction opportunity.
Such action space design leads to large computational costs in high-load environments.
Therefore, the alternative scaling-based approach recomputes only a single scaling factor per timestamp shared with all advertisers and is used in the following algorithms.
DRLB~\cite{wu2018budgetDRLB} defines bid through linear bidding heuristic~(\ref{eq:linear}), where scaling factor $\alpha$ is tuned with RL method. 
FAB~\cite{liu2020dynamicFAB} computes a base bid from predicted CTR and expected CPC and multiplies it by the trained scaling factor in a risk-aware manner. 
Scaling-based approaches naturally support uniform budget pacing (see Section~\ref{subsec::metrics_budget}) since scaling factors directly control budget changes.
USCB~\cite{USCB} replaces PID controller with RL-based algorithms to update coefficients in optimal formulas similar to~(\ref{eq:pid2019}).
Thus, action space based on explicit bids provides finer granularity of bidding process and allows real-time tracking of objectives and constraints.
At the same time, scaling-based action space requires less computational resources and naturally supports budget pacing.


\paragraph{Transition probabilities}
The update of the agent state from $\mathcal{S}$ given an action from $\mathcal{A}$ is controlled by transition probabilities.  
These probabilities affect how agents observe auction dynamics.
Model-free approaches (DRLB~\cite{wu2018budgetDRLB}, FAB~\cite{liu2020dynamicFAB}, USCB~\cite{USCB}) learn policies $\pi$ directly from experience without explicitly modeling transition probability distribution $\mathcal{P}$. 
In contrast, model-based methods explicitly approximate~$\mathcal{P}$ and utilize estimated probabilities in policy learning. 
For example, in RLB~\cite{cai2017realRLB}, $\mathcal{P}$ is represented by market price distributions, while CBRL~\cite{wang2022roiCBRL} approximates transition probabilities with cumulative statistics (e.g., budget consumption, delivery metrics) and market dynamics (e.g., market prices, costs).
Thus, model-based methods offer efficient policy learning strategies with less data by leveraging market structure, while the model's misspecification for $\mathcal{P}$ may lead to convergence issues.
At the same time, model-free methods do not require a model for $\mathcal{P}$; however, training a policy model is slower and requires more data.   

\paragraph{Reward engineering}
The design of the reward function $r$ can take into account both the performance of autobidding algorithms, e.g., ~(\ref{eq::clicks_sum}),~(\ref{eq::conversion_metric}), and cost constraints~(\ref{eq:b_constr}),~(\ref{eq::cost_constraints}).
Basic approaches construct a reward function based solely on the algorithm's performance. 
For example, RLB~\cite{cai2017realRLB} uses click-through rates as rewards, while rewards in CMDP~\cite{du2017improvingCMDP} are total number of clicks~(\ref{eq::clicks_sum}). 
The reward function in DRLB~\cite{wu2018budgetDRLB} is the total number of displays.
FAB~\cite{liu2020dynamicFAB} designs a reward based on a heuristic baseline and provides positive rewards if the number of clicks exceeds the clicks from the baseline, and negative rewards, otherwise.
More advanced approaches incorporate cost constraints in the reward design.
In particular, rewards in USCB~\cite{USCB} are the difference between the total number of displays and the penalty term for violating cost constraints.
In CBRL~\cite{wang2022roiCBRL}, the reward is designed as cumulative delivery (displays or conversions) achieved by the seller, minus a penalty for constraint violations. 
Thus, although performance-based rewards require additional mechanisms to handle constraints, they focus on single objectives, which makes training faster.
In contrast, explicit incorporation of cost constraints in the reward function balances performance and feasibility requirements; however, it requires careful design of penalty terms.

\subsubsection{What metrics are used to evaluate RL autobidding algorithms?}
Evaluation metrics vary significantly across RL autobidding algorithms.
Autobidding algorithms balance an ad's value, e.g., induced clicks or conversions, while satisfying budget constraints ~(\ref{eq:b_constr}) or cost constraints ~(\ref{eq::cost_constraints}). 
These constraints shape both the design of reward functions and the evaluation metrics.
The evaluation metrics based on the constraints measure whether the constraints hold.
The considered RL autobidding algorithms are evaluated in source papers based on the budget constraint, as agents naturally operate with a limited budget. 
RLB~\cite{cai2017realRLB}, DRLB~\cite{wu2018budgetDRLB}, CMDP~\cite{du2017improvingCMDP}, FAB~\cite{liu2020dynamicFAB} focus only on the budget constraint and ignore other cost-related constraints. 
In contrast, USCB~\cite{USCB} and CBRL~\cite{wang2022roiCBRL} extend the budget constraints with cost-related ones, e.g., CPC and CPA~ (\ref{eq::cost_constraints}), etc., and introduce the corresponding metrics.

The ad's value is evaluated using different approaches.
For example, RLB~\cite{cai2017realRLB} and FAB~\cite{liu2020dynamicFAB} report standard advertising metrics such as the number of clicks~(\ref{eq::clicks_sum}), $eCPM$~(\ref{eq::ecpm}), and $eCPC$~(\ref{eq::ecpc}), etc. 
DRLB~\cite{wu2018budgetDRLB} compares the obtained ad's value with the optimal ones computed via a discrete greedy optimization framework~\cite{dantzig1957discrete}.
CMDP~\cite{du2017improvingCMDP} uses total number of clicks~(\ref{eq::clicks_sum}) as the primary metric.
Other methods develop custom evaluation frameworks. 
For example, USCB~\cite{USCB} introduces a custom metric, which combines reward efficiency with constraint violation penalties.
CBRL~\cite{wang2022roiCBRL} combines USCB and CMDP approaches to define evaluation metrics.
This diversity makes systematic comparison of RL autobidding algorithms especially challenging and limits the reproducibility of research findings. 
Table~\ref{tab::rl_models} in Appendix~\ref{app::rl_table} provides a comprehensive comparison of RL autobidding algorithms in terms of the basic RL method, considering constraints and evaluation metrics.


\section{Numerical experiments}

We conducted four experiments in our environment. 
Each of them had the goal of training (or optimizing the hyperparameters) of the algorithm to maximize the metrics $AWR$, $Clicks$, or conversion $Cnv$ in the first three experiments, and to minimize the $RMSE$ in the last one. 
In addition to the target metric, the remaining six metrics were also measured to create a comprehensive picture of the algorithm's performance.
We evaluate classic and RL autobidding algorithms separately in the following subsections.

\subsection{Non-RL autobidding algorithms results}
\label{sec::non_rl_results}
Here, we demonstrate how the classical autobidding algorithms perform in terms of complete set metrics when tuned to the selected target metric. 
We consider $AWR$, the total number of clicks, the total number of conversions, and $RMSE$.

\subsubsection{Tuned for AWR}
\begin{figure}[!ht]
    \centering
    \includegraphics[width=0.5\textwidth]{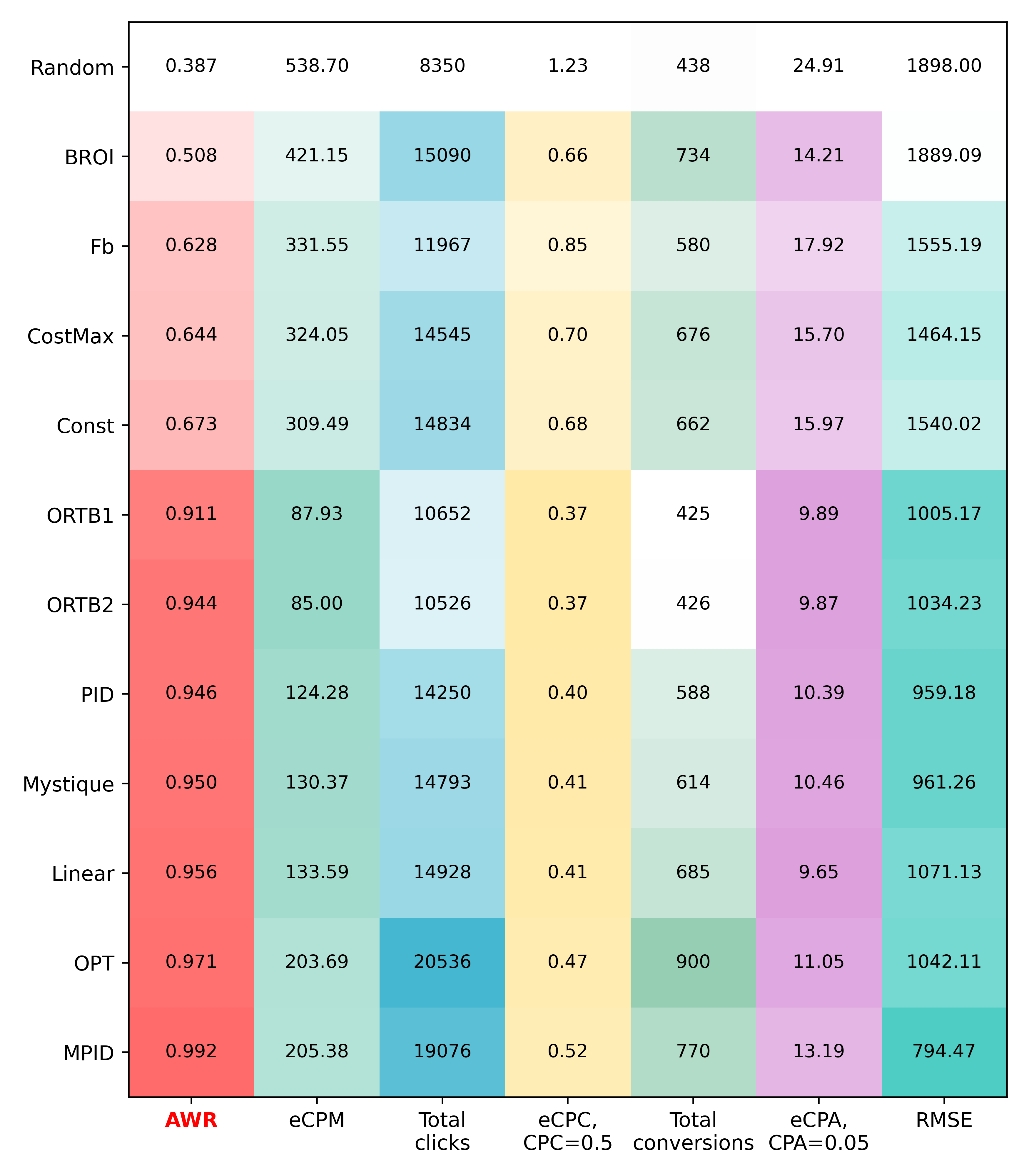}
    \caption{The performance of the classical algorithms tunes on the $AWR$ metric. MPID shows the largest $AWR$ and the smallest $RMSE$ while underperform in terms of other metrics.}
    \label{fig:res_awr}
\end{figure}

The results of the experiment, where algorithms are ranked by the auction win rate~(\ref{eq::awr}), are presented on Figure~\ref{fig:res_awr}. 
OPT~(\ref{eq:pid2019}) and MPID~\cite{yang2019bid} algorithms showed the best $AWR$, while also obtaining a large number of clicks.
Heuristic ORTB approaches~\cite{zhang2014optimal} perform well while tuned on this metric. 
This observation aligns with their design based on the winning rate.
However, tuning on the winning rate makes them less competitive with the total clicks metric.
Notably, tuning for $AWR$ leads to smaller $eCPM$~(\ref{eq::ecpm}) and makes budget pacing uniform.

\subsubsection{Tuned for clicks}
\begin{figure}[!ht]
    \centering
\includegraphics[width=0.5\textwidth]{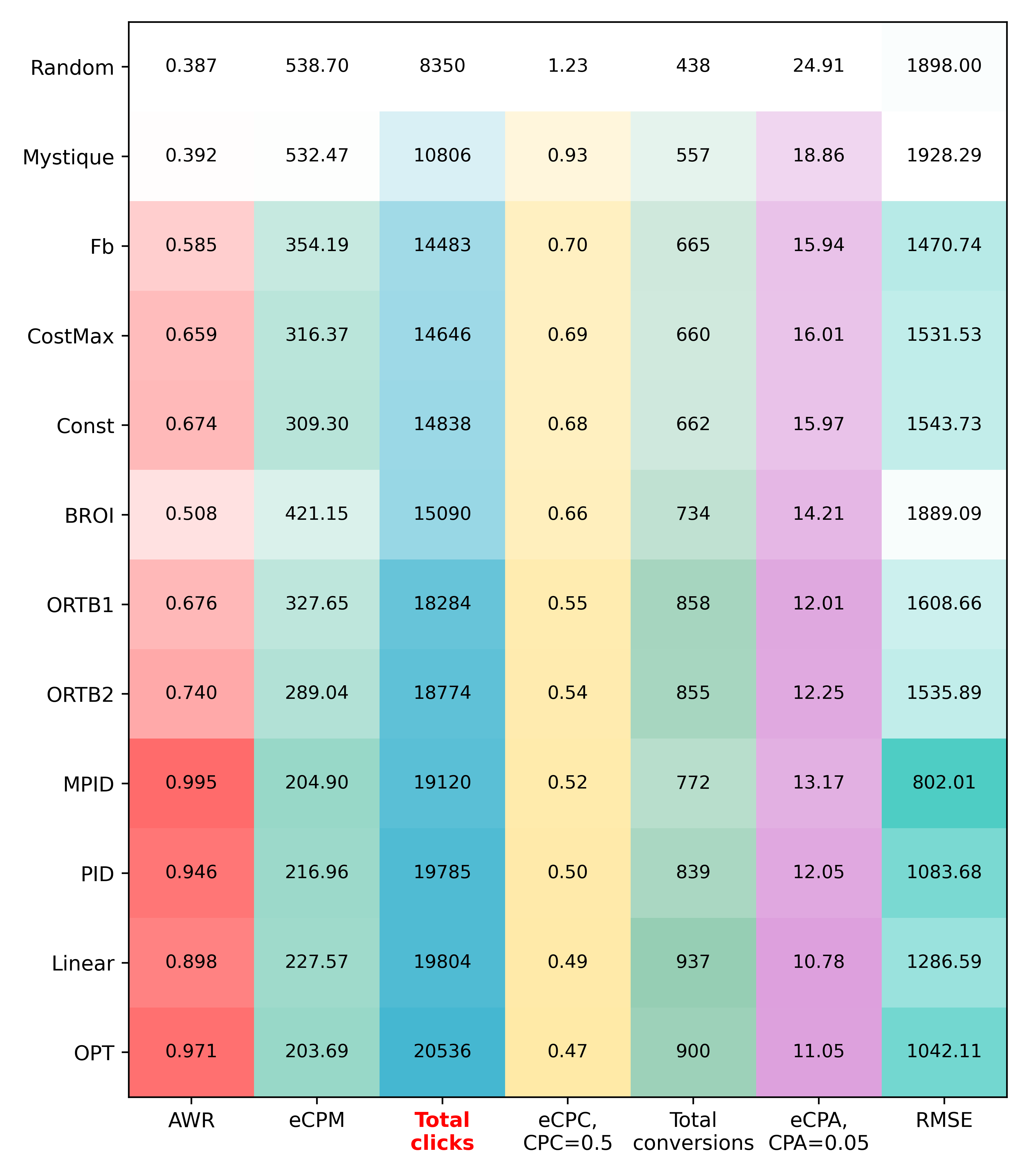}
    \caption{The performance of the classical algorithms tuned on the total number of clicks. Linear and OPT show the largest number of clicks and the smallest eCPC, while underperforming in terms of other metrics.}
    \label{fig:res_clicks}
\end{figure}

Figure~\ref{fig:res_clicks} shows the results of the evaluation of non-RL algorithms, while tuned for the total number of clicks. 
Linear and PID/MPID algorithms show the most promising results. 
Although ORTBs are less effective, their performance is still competitive.
A notable observation is the inverse correlation between $eCPC$ and $Clicks$, indicating that higher click volumes are associated with a lower cost per click. 
It follows that algorithms excelling in $Clicks$ are likely to perform well on most other metrics, with $RMSE$ being a notable exception.

\subsubsection{Tuned for conversion}
\begin{figure}[!ht]
    \centering
    \includegraphics[width=0.5\textwidth]{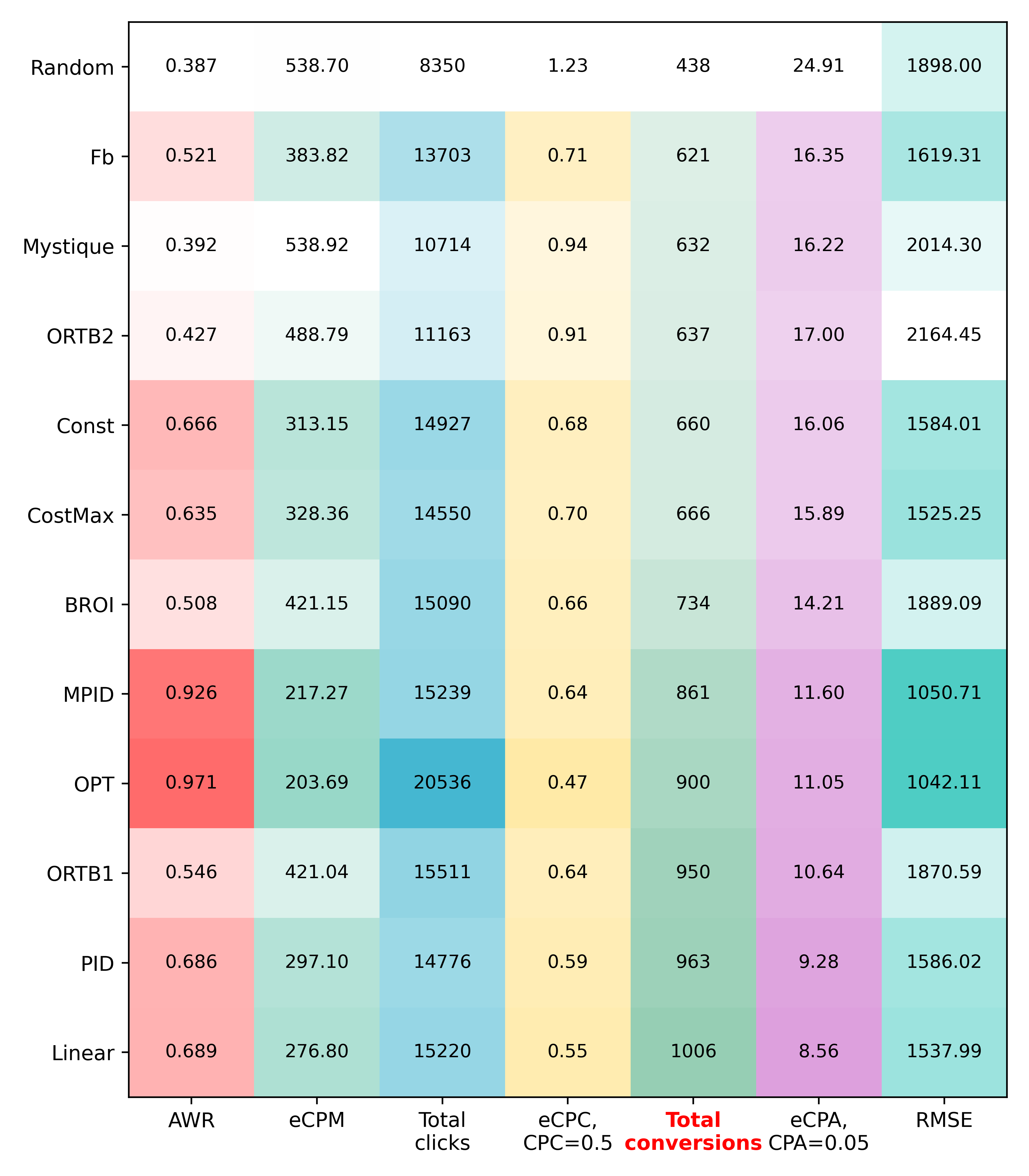}
    \caption{The performance of the classical algorithms tuned on the total number of conversions. Linear and PID show the largest number of conversions, while underperforming in terms of other metrics.}
    \label{fig:res_conversions}
\end{figure}

The results of the algorithm comparison with tuning on the Total Conversion metric~(\ref{eq::conversion_metric}) are presented in Figure~\ref{fig:res_conversions}.
Linear~\cite{zhang2014optimal} and PID models show the best performance.
Notably, conversion optimization does not result in a higher auction win rate.
As expected, conversion optimization yields the smaller $eCPA$ compared to click optimization. 

\subsubsection{Tuned for RMSE}
\begin{figure}[ht]
    \centering
    \includegraphics[width=0.5\textwidth]{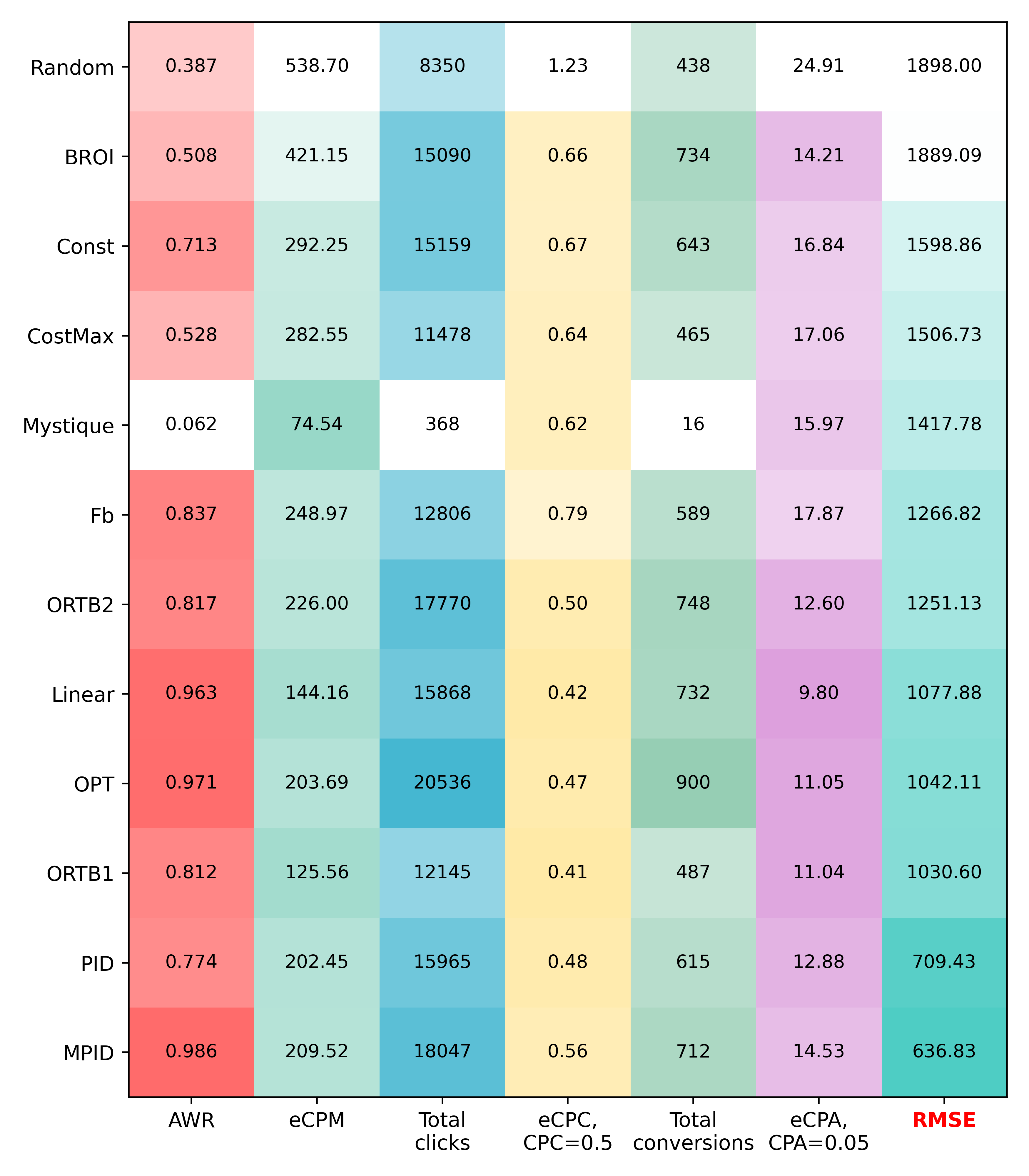}
    \caption{The performance of the classical algorithms tuned on the $RMSE$ metric. MPID and PID give the smallest $RMSE$ while underperforming in terms of other metrics.}
    \label{fig:res_rmse}
\end{figure}

PID and MPID controllers~\cite {yang2019bid}, which address the budget pacing problem, show the best results, see Figure~\ref{fig:res_rmse}.
However, two other controllers, Fb~\cite{zhang2016feedback} and Mystique~\cite{stram2024mystique}, show poor performance. 
In most cases, tuning for RMSE allows other metrics to remain competitive.

\subsection{RL autobidding algorithms results}
We consider RLB, DRLB, USCB, and CBRL autobidding algorithms and learn them to maximize the total number of clicks.

Hyperparameters were taken from the original publications if specified and manually tuned otherwise. Table~\ref{tab:algorithm_results} shows that CBRL gives the uniformly best metrics except total conversions $Cnv$, which aligns with its advanced design.
RLB provides the best total conversion~$Cnv$, despite being a classical RL algorithm.
Finally, DRLB and USCB provide intermediate performance since they depend on many unknown hyperparameters that require careful tuning.
Thus, comparison of results from Section~\ref{sec::non_rl_results} indicates that RL autobidding algorithms underperform the well-tuned classical algorithms and their deployment is challenging. 

\begin{table}[!h]
\centering
\caption{Summary on the performance metrics for the considered RL autobidding algorithms. The best values in every column are bold, and the second-best ones are underlined. CBRL uniformly outperforms other methods in our setup.}
\label{tab:algorithm_results}
\begin{tabular}{lrrrrrrr}
\toprule
 & $AWR$ & $eCPM$ & $Clicks$ & $eCPC$ & $Cnv$ & $eCPA$ & $RMSE$ \\
\midrule
RLB       & 0.41 & 513.82 & 12393.50 & 0.83 & \textbf{681.90} & 16.25 & 2187.48 \\
DRLB      & \underline{0.44} & 478.39 & \underline{12795.25} & \underline{0.79} & 663.57 & \underline{16.17} & 2195.81 \\
USCB      & 0.42 & \underline{194.80} & 5022.08  & \underline{0.79} & 217.71 & 19.11 & \underline{1382.32} \\
CBRL      & \textbf{0.96} & \textbf{139.95} & \textbf{15553.92} & \textbf{0.42} & \underline{667.04} & \textbf{10.32} & \textbf{960.89} \\
\bottomrule
\end{tabular}
\end{table}

\section{Conclusion}   

We have presented the Autobidding Arena framework for evaluating a wide range of classical and RL autobidding algorithms.
To evaluate algorithms, we have designed the environment using a publicly available dataset.
We have chosen the most common metrics, which are transparent and platform-free, from an extensive collection of autobidding papers.
According to these metrics, we demonstrate the performance of the selected classical and RL-based autobidding algorithms in our simulation environment.
We empirically confirm that tuning hyperparameters in autobidding algorithms based on a predefined target metric can degrade other metrics.
We observed that well-tuned classical algorithms outperform RL algorithms running with basic hyperparameters. 

\bibliographystyle{unsrt}
\bibliography{lib}

\appendix

\section{Generation of CTR and CVR from a $p$-value}\label{appendix:CTRCVR}

Bids are placed for the entire time step at once, but are personalized: for each item, a unique value $pValue_{as}$ and $wp_a$ are stored in the dataset for each auction $a$. 
$pValue_{as}$ is the probability of conversion given an impression, which in our terminology means $pValue_{as} = CTR_{as} \cdot CVR_{as}$.

Thus, the separation of $pValue_{as}$ into independent $CTR_{as}$ and $CVR_{as}$ values for the item in each auction is needed.

To synthesize realistic click-through rate (CTR) and conversion rate (CVR) signals for the benchmark, we decomposed a given scalar $p$-value into two stochastic components. The process adheres to three fundamental requirements. First, the product equals $p$. Second, the mean ratio is approximately $2:1$ which coincides with the ratio on other huge open-sourced benchmarks \cite{khirianova2025bat}. Finally, non-degenerate randomness is injected to avoid a purely deterministic factorization.

We introduce an auxiliary scalar $a$ and target means $CTR_0=2a,\; CVR_0=a$. From the noiseless constraint $CTR_0\cdot CVR_0=p$ we obtain $a=\sqrt{\frac{p}{2}}.$
To model variability, we add independent Gaussian perturbations $\varepsilon$ and~$\beta$ with zero mean and variance proportional to $p$, namely $\varepsilon,\beta\sim\mathcal{N}(0,p/8)$. The observed metrics are defined as $\text{CTR} = 2a - \varepsilon$ and $\text{CVR} = a + \beta$. Imposing the exact product constraint CTR$\cdot$ CVR $= p$ yields a quadratic equation in $a$; choosing the positive root gives the closed form
\begin{equation*}
a \;=\; \frac{1}{4}\Bigl(\varepsilon - 2\beta + \sqrt{(\varepsilon + 2\beta)^2 + 8p}\Bigr).
\end{equation*}
Finally, CTR and CVR are obtained by substituting this $a$ into the definitions above. In practice we clip the resulting probabilities to a small numerical-safe interval (e.g. $[10^{-6},\,1-10^{-6}]$).

This procedure preserves the invariant $\mathrm{CTR} \cdot \mathrm{CVR} = p$ by construction, yields an expected ratio close to $2:1$, and introduces realistic sample-wise variability.

\section{RL autobidding algorithms comparison}
\label{app::rl_table}
To demonstrate metrics and algorithms types diversity we compared them in Table~\ref{tab::rl_models}. 
This table shows the diversity of algorithms evaluation in the prior works. 

\begin{table*}[!h]
\centering
\caption{Comparison of RL methods for autobidding. 
Metrics Types categorize the authors evaluation approaches: performance metrics (TC (Total Clicks), TI (Total Impressions)) measure user engagement; 
cost-efficiency metrics (CPC, CPA, CPM, ROI) evaluate economic effectiveness; 
budget utilization (ConBdg) tracks spending efficiency; 
reward-oriented metrics (RO) assess algorithm performance against theoretical bounds or baselines through regret analysis and optimization ratios. 
CPx means both CPA and CPC constraints from~\ref{eq::cost_constraints}; B refers to the budget constraint~\ref{eq:b_constr}.
A, C, D criteria are listed in Section~\ref{app:algos} of Appendix.
Bold model name means that the model participated in our comparison, see Table~\ref{tab:algorithm_results}.
}
\resizebox{\linewidth}{!}{
\begin{tabular}{cccccc}
  \toprule
  Method name & Base RL algorithm & Key features & Constraints & Metrics Types & \textbf{ACD}\\
  \midrule
  \textbf{RLB}~\cite{cai2017realRLB} & Value Iteration~\cite{sutton1998reinforcement} & Impression-level MDP, model-based & B & TC, CPM, CPC & \checkmark\checkmark\checkmark \\
  LADDER~\cite{wang2017ladder} & DQN & Asynchronous DQN & B+CPC & Revenue, ROI  & \checkmark\texttimes\checkmark \\
  CMDP~\cite{du2017improvingCMDP} & Batch RL~\cite{lange2012batch} & Constrained MDP formulation & B &  TC, eCPC  & \checkmark\texttimes\checkmark \\
  \textbf{DRLB}~\cite{wu2018budgetDRLB} & DQN~\cite{mnih2015humanDQN} & Model-free, bid factor control & B & TC, RO  & \checkmark\checkmark\checkmark\\
  RMDP~\cite{zhao2018deepRMDP} & DQN~\cite{mnih2015humanDQN} & Sponsored Search RTB & B & ROI  & \checkmark\texttimes\checkmark \\
  ClusterA3C~\cite{clusterA3C} & A3C~\cite{mnih2016asynchronousA3C} & KPI curve clustering & B& RO  & \checkmark\texttimes\texttimes \\ 
  FAB~\cite{liu2020dynamicFAB} & TD3~\cite{pmlr-v80-fujimoto18aTD3} & Continuous action space & B & TC, TI, CPC & \checkmark\checkmark\checkmark\\
  \textbf{USCB}~\cite{USCB} & DDPG~\cite{lillicrap2015continuousDDPG} & Multiple Constraint, model-free, off-policy & B+CPx  &  RO  & \checkmark\checkmark\checkmark\\
  \textbf{CBRL}~\cite{wang2022roiCBRL} & SAC & Bayesian Q-learning, & B+CPx & RO, ROI & \checkmark\checkmark\checkmark \\
  &&Curriculum learning, Posterior sampling&&\\
  SORL~\cite{mou2022sustainableSORL} & DDPG~\cite{lillicrap2015continuousDDPG} & Online-offline inconsistency fix & B & TI, ROI, CPA, ConBdg & \checkmark\texttimes\checkmark\\
  PerBid~\cite{zhang2023personalizedPerBid} & CaDM~\cite{lee2020context}& Fairness investigation & B+CPC & RO, Gini Coefficient & \checkmark\texttimes\checkmark\\
  \bottomrule
\end{tabular}
}
\label{tab::rl_models}
\end{table*}

\section{Performance Metrics Evaluation}\label{appendix:MetrEval}

The \textbf{lack of standardization} in performance metrics for auto-bidding constitutes a fundamental issue, as comparisons between new algorithms and baselines or evaluations of their effectiveness are frequently based on non-transparent or non-universal metrics.

Poorly designed metrics can be divided into categories:

\begin{itemize}
    \item metrics \textbf{irrelevant to seller-side auto-bidding:} social welfare metrics, Gini Coefficient, and platform revenue metrics that primarily serve the platform's interests, as used in \cite{liu2023auto, balseiro2021robust,jeunen2023off, finster2023competitive, zhang2023personalized}. Moreover, there exists a variety of game theory metrics - liquid welfare \cite{lucier2024autobidders}, transferable welfare \cite{babaioff2020non}, etc., but they exploit the concept of multi agent games and corresponds for the optimal policy for each agent in the equilibrium, while this paper considers the approach of only one bidder interacting with the system;
    
    \item \textbf{production-specific or business-specific} metrics with limited generalizability for researchers: engagement rate \cite{geyik2016joint}, business metrics like shopping cart addition volume \cite{guan2021multi}, GMV \cite{guan2021multi, zhu2017optimized,zhang2014optimal}, and its derivatives such as RPM \cite{zhu2017optimized} and Rscore \cite{zhang2014optimal}, advertising effect \cite{maehara2018optimal}, the difference between actual delivery and best possible delivery \cite{wang2022roi};
    
    \item \textbf{non-intuitive derivative} metrics employing custom normalizations or formulaic extensions that overcomplicate comparisons: Click-to-Spend Yield Ratio (CYR) \cite{kong2022not}, CTR lift \cite{wang2017ladder}, logarithm of CPX and revenue-cost ratio with complex normalization \cite{zhou2021efficient}. Here we also include metrics \textbf{losing relevance} due to flawed design: e.g., the CPCratio formula in \cite{yang2019bid, zhang2016feedback, kitts2017ad}, which ignores constraint violations exceeding 10\%, consequently always equaling 1 in the paper's comparison tables.
\end{itemize}
Having excluded all the above categories from consideration, we selected only transparent universal autobidding metrics.

\begin{table}[!h]
\centering
\caption{Evaluation of auto-bidding performance metrics based on selection criteria}
\label{tab:metrics_evaluation}
\begin{tabular}{p{6cm}ccc}
\toprule
Metric & 
Relevance & 
Universality & 
Transparency \\
\midrule
\textbf{Auction Win Rate (AWR)} \cite{gligorijevic2020bid} & \checkmark & \checkmark & \checkmark \\
\textbf{Total Clicks} \cite{zhang2014optimal} & \checkmark & \checkmark & \checkmark \\
\textbf{Total Conversions} \cite{yang2019bid} & \checkmark & \checkmark & \checkmark \\
\textbf{Expected Cost Per Mille (eCPM)} \cite{cai2017real} & \checkmark & \checkmark & \checkmark \\
\textbf{Expected Cost Per Click (eCPC)} \cite{gligorijevic2020bid} & \checkmark & \checkmark & \checkmark \\
\textbf{Expected Cost Per Action (eCPA)} \cite{zhou2021efficient} & \checkmark & \checkmark & \checkmark \\
\textbf{Budget pacing (RMSE)} \cite{khirianova2025bat} & \checkmark & \checkmark & \checkmark \\
Social Welfare \cite{balseiro2021robust} & \texttimes & \checkmark & \checkmark \\
Gini Coefficient \cite{zhang2023personalized} & \texttimes & \checkmark & \checkmark \\
Liquid welfare \cite{lucier2024autobidders} & \texttimes & \checkmark & \checkmark \\
Transferrable welfare \cite{babaioff2020non} & \texttimes & \checkmark & \checkmark \\
Engagement Rate \cite{geyik2016joint} & \checkmark & \texttimes & \checkmark \\
Shopping Cart Addition Volume \cite{guan2021multi} & \checkmark & \texttimes & \checkmark \\
GMV \cite{guan2021multi} & \checkmark & \texttimes & \checkmark \\
RPM \cite{zhu2017optimized} & \checkmark & \texttimes & \checkmark \\
Rscore \cite{zhang2014optimal} & \checkmark & \texttimes & \checkmark \\
Advertising Effect \cite{maehara2018optimal} & \checkmark & \texttimes & \checkmark \\
Delivery Difference \cite{wang2022roi} & \checkmark & \texttimes & \checkmark \\
Click-to-Spend Yield Ratio (CYR) \cite{kong2022not} & \checkmark & \checkmark & \texttimes \\
CTR lift \cite{wang2017ladder} & \checkmark & \checkmark & \texttimes \\
CPX version \cite{zhou2021efficient} & \checkmark & \checkmark & \texttimes \\
Revenue-cost ratio \cite{zhou2021efficient}  & \checkmark & \checkmark & \texttimes \\
CPCratio  \cite{yang2019bid} & \checkmark & \checkmark & \texttimes \\
\bottomrule
\end{tabular}
\end{table}

\newpage
\section{Algorithms}\label{app:algos}

Since the criteria for selecting algorithms are their abundance, compatibility, distinctness, modernity and high citation rate, and variability, we evaluated the algorithms as follows.

Tables \ref{table:heuristics}, \ref{table:bp}, \ref{table:constraints}, \ref{table:design}, \ref{table:shading} include articles on non-RL autobidding from sources with a high citation rate since 2014. Next, the algorithms are divided by the purpose of their creation, for clarity of variability: heuristics and simple baselines (Table \ref{table:heuristics}), budget pacing algorithms (Table \ref{table:bp}), algorithms with ROI/CPA/CPC constraints (Table \ref{table:constraints}), auction design oriented algorithms (Table \ref{table:design}), and algorithms with predictors and shading (Table \ref{table:shading}). 

Finally, we designated the remaining three criteria abundance, compatibility, distinctness with the letters "A", "C", "D" respectively and indicated whether each algorithm satisfies the specified property. A check mark in column "A" indicates that the algorithm is fully presented in the paper, without hidden formulas for functions and variable values. A check mark in column "C" indicates that the algorithm is integrated into our proposed environment, meaning the input and output data correspond to the capabilities of the environment. The check mark in column "D" means that the algorithm responsible for actually forming the bid is separate from other parts of the architecture described in the article, for example, the one that predicts CTR or winning price value.

If the algorithm met all the selection criteria, we provided the bid formula in the "Formula" column. Otherwise, we marked the presence of a final bid formula in the original article with a "$+$" sign and its absence with a "$-$" sign.

\begin{table*}[!h]
\caption{Evaluation of heuristic auto-bidding algorithms based on selection criteria}
\centering
\resizebox{\linewidth}{!}{
\begin{tabular}{cccp{5.3cm}ccc}
\toprule
\textbf{Paper} & \textbf{Concept/Algorithm} & \textbf{Year} & \textbf{Formula} & \textbf{A} & \textbf{C} & \textbf{D}\\
\midrule
\cite{zhang2014optimal} & \textbf{Constant} bid & 2014 & $bid_{const} = b_0$ & \checkmark & \checkmark & \checkmark\\
\cite{zhang2014optimal} & \textbf{Linear} bid & 2014 & $bid_{lin} = a \cdot CTR$ & \checkmark & \checkmark & \checkmark\\
\cite{zhang2014optimal} & \textbf{Random} bid & 2014 & $bid_{rand}=random(bid_{\min}, bid_{\max})$ & \checkmark & \checkmark & \checkmark\\
\cite{zhang2014optimal} & \textbf{CostMax} & 2014 & $bid_{CostMax} = b \cdot CPC$ & \checkmark & \checkmark & \checkmark \\
\cite{zhang2014optimal} & Optimal non-linear concave bidding \textbf{ORTB$_{1}$} & 2014 & 
\begin{minipage}[t]{7.8cm}
 $b_{ORTB1}=\sqrt{\frac{c}{\lambda}\,CTR + c^2} - c$
\end{minipage}
 & \checkmark & \checkmark & \checkmark\\
\cite{zhang2014optimal} & Optimal non-linear concave bidding \textbf{ORTB$_{2}$} & 2014 & 
$+$
 & \checkmark & \checkmark & \checkmark\\
 \cite{zhang2017managing}& \textbf{Risk-Based} & 2017 & $bid_{Risk} = CTR-\alpha \cdot CTR_{std}$ &\checkmark & \checkmark & \checkmark\\
\cite{kitts2017ad} & CostMin & 2019 & $+$ & \checkmark & \checkmark & \texttimes \\
\bottomrule
\end{tabular}
}
\label{table:heuristics}
\end{table*}

\begin{table*}[!h]
\caption{Evaluation of budget pacing auto-bidding algorithms based on selection criteria}
\centering
\begin{tabular}{cp{5cm}ccccc}
\toprule
\textbf{Paper} & \textbf{Concept/Algorithm} & \textbf{Year} & \textbf{Formula} & \textbf{A} & \textbf{C} & \textbf{D}\\
\midrule
\cite{chen2023coordinated} & Coordinated Pacing : select representative + adaptive pacing & 2023 & $+$
& \checkmark & \texttimes & \checkmark\\
\cite{chen2023coordinated} & Hybrid Coordinated Pacing & 2023 & $+$
&\checkmark & \texttimes & \checkmark\\
\cite{stram2024mystique} & Soft-throttling pacing \textbf{Mystique} & 2024 & $+$ 
&\checkmark & \checkmark & \checkmark\\
\bottomrule
\end{tabular}
\label{table:bp}
\end{table*}

\begin{table*}[!h]
\caption{Evaluation of auto-bidding algorithms with ROI/CPA/CPC constraints based on selection criteria}
\centering
\begin{tabular}{cp{8.5cm}ccccc}
\toprule
\textbf{Paper} & \textbf{Concept/Algorithm} & \textbf{Year} & \textbf{Formula} & \textbf{A} & \textbf{C} & \textbf{D}\\
\midrule
\cite{zhang2015statistical} & Statistical CPA-CPM arbitrage with wp and CVR approximation & 2015 & $+$ &\texttimes & \texttimes & \texttimes \\
\cite{xu2015smart} & Budget and eCPC pacing through probabilistic throttling & 2015 & $-$ & \checkmark & \texttimes & \checkmark\\
\cite{geyik2016joint} & Joint optimization for mullti-KPI & 2016 &
$+$
& \texttimes & \checkmark & \texttimes \\
\cite{zhang2016feedback} & \textbf{Fb} - PID controller for stabilizing eCPC and AWR; exponential actuator & 2016 & $+$ & \checkmark & \checkmark & \checkmark\\
\cite{zhang2016feedback} & \textbf{FbWL} - PID/WL controllers for stabilizing eCPC and AWR; exponential actuator & 2016 & $+$ &\checkmark & \checkmark & \checkmark\\
\cite{zhu2017optimized} & OCPC: auto-bidding for target CPA/ROI; eCPC redistribution in ranking & 2017 & $+$ & \checkmark & \texttimes & \texttimes\\
\cite{yang2019bid} & Optimal solution \textbf{OPT}, $p,q$ as hyperparameters & 2019 &
$+$
& \checkmark & \checkmark & \checkmark\\
\cite{yang2019bid} & budget+CPC constraint, \textbf{PID} on $p,q$ & 2019 &
$+$
& \checkmark & \checkmark & \checkmark\\
\cite{yang2019aiads} & Target-CPA automated bidding with multiplicative factors; end-to-end targeting/creation system & 2019 & $-$ & \texttimes & \checkmark & \texttimes \\
\cite{yang2019bid} & Budget+CPC constraint with multivariable control, \textbf{MPID} on $p,q$ & 2019 &
$+$
& \checkmark & \checkmark & \checkmark\\
\cite{liu2023auto} & Budget+ROI; optimal shading; truthfulness; with ROI and utility approximation & 2023 &
$+$ & \checkmark & \texttimes & \texttimes \\
\cite{BiCB} & Optimal solution \textbf{OPT}, $p,q$ as hyperparameters with LGBM prediction on Cost and CTR& 2025 &
$+$
& \checkmark & \checkmark & \checkmark\\
\bottomrule
\end{tabular}
\label{table:constraints}
\end{table*}

\begin{table*}[!h]
\caption{Evaluation of auction design oriented auto-bidding algorithms based on selection criteria}
\centering
\begin{tabular}{cp{7cm}ccccc}
\toprule
\textbf{Paper} & \textbf{Concept/Algorithm} & \textbf{Year} & \textbf{Formula} & \textbf{A} & \textbf{C} & \textbf{D}\\
\midrule
\cite{balseiro2015repeated} & Reserves and boosts improve welfare and revenue & 2021 & 
$-$ & \checkmark & \checkmark & \texttimes \\
\cite{chen2023coordinated} & Individual adaptive pacing: budget-smoothed shading \textbf{CB} & 2023 & $ bid_{CB} = \frac{a\cdot CTR_t \cdot CVR_t}{1+\lambda}$ &\checkmark & \checkmark & \checkmark \\
\cite{lucier2024autobidders}& optimizing welfare, budget and ROI constraints \textbf{BROI} &2024& $bid_{BROI} = \frac{CTR_{t} CVR_t}{1 + \mu_{t}}$ &\checkmark & \checkmark & \checkmark\\
\cite{su2024a} & RTB environment benchmark (GSP, multi-slot), data generation, baseline algorithms. & 2024 & $-$ &\checkmark & \texttimes & \texttimes \\
\bottomrule
\end{tabular}
\label{table:design}
\end{table*}

\begin{table*}[!h]
\caption{Evaluation of auto-bidding algorithms with predictors and shading based on selection criteria}
\centering
\begin{tabular}{lp{8cm}ccccc}
\toprule
\textbf{Paper} & \textbf{Concept/Algorithm} & \textbf{Year} & \textbf{Formula} & \textbf{A} & \textbf{C} & \textbf{D}\\
\midrule
\cite{lin2016combining} & Threshold strategy by efficiency: CTR and predicted WP  & 2016 & $+$ &\checkmark & \texttimes & \texttimes \\
\cite{gligorijevic2020ciKM} & Shading in 1st price: FM model for shading factor & 2020 & $+$ & \checkmark & \texttimes & \checkmark\\
\cite{pan2020adkdd} & Win-Rate shading: logistic $\mathbb{P}(win \mid bid)$ & 2020 & $+$ & \checkmark & \texttimes & \checkmark\\
\cite{zhou2021efficient} & Estimating min-win price distribution + golden-section for $bid$ & 2021 & $+$ & \checkmark & \texttimes & \checkmark\\
\bottomrule
\end{tabular}
\label{table:shading}
\end{table*}

\end{document}